\documentclass[12pt]{article}
\newtheorem{lemma}{Lemma}[section]

\newtheorem{theorem}{Theorem}[section]

\newtheorem{definition}{Definition}[section]

\def\QED{\quad q.e.d.} 
\def\qed{\QED}
\def\mathbb#1{#1}
\def\IR{{\rm I\!R}} 

\def\zed{{\mathchoice {\hbox{$\sf\textstyle Z\kern-0.4em Z$}}
{\hbox{$\sf\textstyle Z\kern-0.4em Z$}}
{\hbox{$\sf\scriptstyle Z\kern-0.3em Z$}}
{\hbox{$\sf\scriptscriptstyle Z\kern-0.2em Z$}}}
}
\def\zedd{\zed^{\!d}}
\def\reff#1{(\ref{#1})}

\def\LL#1{\label{#1}
}

\def\KX#1{\hbox{\kern#1ex}}

\def\RF#1{\KX{0.10}\ref{#1}\KX{0.10}%
}

\def\BAR%
{\hbox{\kern0.9ex}\raisebox{-0.9ex}{\rule{0.15ex}{3.0ex}}\hbox{\kern0.9ex}}

\def\HB#1{\hbox{\kern1ex #1 \kern1ex}}
\def\FORM#1{\begin{equation}#1\end{equation}}

\def\SET#1{\{#1\}}
\def\IMPLIES{~\Longrightarrow~}
\def\CONST{\hbox{\rm const}}
\def\BRA#1{ \left( #1 \right) }
\def\NEG#1{\neg #1}

\def\M{{\cal M}}
\def\epsi{\varepsilon}

\def\major{\hbox{\rm major}}
\def\PHI{\overline{\phi}}
\def\ind{\hbox{\rm Ind}}
\def\conf{\hbox{\rm Conf}}
\def\scalar#1{\langle #1 \rangle}
\def\one{\hbox{\bf 1}}

\begin{document}
\sloppy
\bibliographystyle{plain}

\vspace*{40pt}

\begin{center}
  \large\bf
  Non-Gibbsianness of the invariant measures \\
  of non-reversible cellular automata with totally asymmetric noise\\

\vspace*{20pt}

Roberto Fern\'andez%
\footnote{University of Rouen, France. 
E-mail Roberto.Fernandez@univ-rouen.fr.}
and Andr\'e Toom%
\footnote{UFPE, Brazil. 
E-mail toom@bernoulli.de.ufpe.br or toom@member.amd.org.
Supported by FAPESP, grant \# 98/15994-0 
and CNPq, grant \# 300991/1998-3.} 

\end{center}

\begin{abstract}
  
  We present a class of random cellular automata with multiple
  invariant measures which are all non-Gibbsian.  The automata have
  configuration space $\SET{0,1}^{\zedd}$, with $d > 1$, and they
  are noisy versions of automata with the ``eroder property''.  The
  noise is totally asymmetric in the sense that it allows random
  flippings of ``0'' into ``1'' but not the converse.  We prove that
  all invariant measures assign to the event ``a sphere with a large
  radius $L$ is filled with ones'' a probability $\mu_L$ that is too
  large for the measure to be Gibbsian. For example, for the NEC
  automaton $(-\ln \mu_L) \asymp L$ while for any Gibbs measure the
  corresponding value is $\asymp L^2$.

\end{abstract}

{\bf Key words:~} Gibbs vs. non-Gibbs measures, cellular automata, 
invariant measures, non-ergodicity, eroders, convex sets.

\newpage
\large\baselineskip 30pt

\vspace*{40pt}

\section{Introduction}
Studies of cellular automata and of their continuous-time counterpart,
the spin-flip dynamics, have been successful in determining how
many invariant measures the automaton or dynamics have.  Much less is
known about properties of these measures.  A natural question is
whether they are Gibbsian, that is whether they could correspond to
measures describing the equilibrium state of some statistical
mechanical system.  There are two categories of evolutions ---both
with local and strictly positive updating rates--- for which the
answer is known to be positive: (1) If the updating prescription has a
high level of stochasticity ---\emph{high noise regime}---, in which
case Gibbsianness comes together with uniqueness of the invariant
measure \cite{lebmaespe90,maevel93,lormae97}; and (2) if the updating
satisfies a detailed balance condition for some Boltzmann-Gibbs
weights \cite{maevel94}.  Known cases of non-Gibbsianness, on the
other hand, refer to automata where the updating rates are either
non-strictly positive \cite{lebsch88}, \cite[Chapter 7]{vel95} or
non-local \cite{marsco91}.

In this paper we present some examples of stochastic
\emph{non-reversible automata} ---that is, automata not satisfying 
any form of detailed balance---, with multiple invariant measures, 
all of them non-Gibbsian.  Our class of automata can be seen as a
generalization of the North-East-Center (NEC) majority model
introduced in \cite{petrovskaya} and discussed in many papers.  
Its non-ergodicity was first proved in \cite{too74} (see also the
discussion in \cite{lebmaespe90}) and later by another method in
\cite{BraGra}.  Also it was simulated more than once
\cite{BenGri,mak97,mak99}.  Models of this sort are obtained by 
superimposing stochastic errors (noise) to deterministic automata 
having the so-called \emph{eroder property}: finite islands of 
aligned spins, within a sea of spins aligned in the opposite direction,
disappear in a finite time.

We allow only \emph{one-sided} noise or stochastic error ---a ``0''
can stochastically be turned into a ``1'', but not the reverse.  Thus
some of our transition rates are zeros and therefore the ``dichotomy''
result of \cite[Corollary 1]{maevel94} is not applicable.  Our work
does not settle the long-standing issue of the Gibbsianness of the
invariant measures of NEC models with non totally asymmetric noise.
There are conflicting arguments and evidences for the model with
symmetric noise: An interesting heuristic argument has been put
forward \cite[Chapter 5]{vel95} pointing in the direction of
Gibbsianness, and a couple of pioneer numerical studies yielded
findings respectively consistent with Gibbsianness \cite{mak97} and
non-Gibbsianness \cite{mak99}.  However, we hope that the simple
non-Gibbsianness mechanism clearly illustrated by our examples could
be a useful guide and reference for the study of the more involved
two-way-noise situation.

In our examples, non-Gibbsianness shows up in the same way as in 
the basic voter model \cite{lebsch88}: Large droplets of aligned
(``unanimous'') spins have too large probability for the invariant
measures to be Gibbsian.  More precisely, we show that once a 
suitable ``spider'' of ``1'' appears, the dynamics causes the
alignment of the spins in a neighboring sphere.  This sort of
damage-spreading property (or error-correcting deficiency) implies
that the presence of a sphere of ``1'' is penalized by the invariant
measures only as a sub-volume exponential.  This contradicts well
known Gibbsian properties.  In fact, we can be more precise.  Gibbsian
measures are characterized by two properties \cite{koz74}: uniform
non-nullness and quasilocality.  As we comment in Section \ref{s.ent},
the large probability of aligned droplets means that the invariant
measures can not be uniformly non-null.  More generally, such
invariant measures can not be the result of block renormalizations of
non-null, in particular Gibbsian, measures.  Furthermore, known
arguments \cite{entlor96} (briefly reviewed in Section \ref{s.ent}
below), imply that if one of these measures is not a product measure,
then its non-Gibbsianness is preserved by further single-site
renormalization transformations.

\section{Simple examples}
\label{s.simple}

Before plunging into the technical and notational details needed to
describe our results in full generality, we would like to present some
simple examples that contain the essential ideas.  The examples are
defined on the configuration space $\SET{0,1}^{\zed ^2}$.

\paragraph{Example 1:  The NEC model}.
Its deterministic version is defined by a translation-invariant
parallel updating defined by the rule
\begin{equation}
  x^{t+1}_{\rm det}(0,0) \;=\; 
  \major\Bigl\{x^t(0,1)\,,\, x^t(1,0)\,,\, x^t(0,0)\Bigr\}\;,
  \label{r.2}
\end{equation}
where $x^t(i,j)$ denotes the configuration at site $(i,j)\in \zed ^2$
immediately after the $t$-th iteration of the transformation and
$\major : \SET{0,1}^{2k+1} \to \SET{0,1}$ is the majority 
function, i.e. the Boolean function of any odd number of arguments,
which equals ``1'' if and only if most of its arguments equal ``1''.
This prescription yields an evolution, which is symmetric with respect
to the flip $0\leftrightarrow 1$ [a function with this property is
called a \emph{self-spin-flip function} in Section \ref{s.gen} below].
We consider a noisy version, where in addition spins ``0'' flip into
``1'' independently with a certain probability $\epsi$, while spins
``1'' remain unaltered. This corresponds to an stochastic updating
\begin{equation}
  \label{eq:r1}
  {\rm Prob}\Bigl(x^{t+1}(i,j)=0\,\Bigm|\,x^t\Bigr) \;=\; 
(1-\epsi)\,\Bigl[1-x^{t+1}_{\rm det}(i,j)\Bigr]\;.
\end{equation}
The ``all-ones'' delta-measure $\delta_1$ is invariant for this automaton.  
For small $\epsi$ there is at least another invariant
measure (see lemma \ref{l.nonuniq} below).

Let us start with the following simple observations which are
immediate consequences of the NEC rule \reff{r.2} and the
one-sidedness of the noise:
\begin{itemize}
\item[(i)] Horizontal lines (parallel to axis $i$) filled with
  spins ``1'' remain invariant under the evolution.
\item[(ii)] The same invariance holds for vertical lines
  (parallel to axis $j$) filled with spins ``1''.
\item[(iii)] After one evolution-step (that is, after one parallel
  updating of all the spins), a line of slope $-1$ filled with spins
  ``1'' moves into the parallel line immediately to the South-West.
\item[(iv)] If the (infinite) ``spider'' formed by the $i$-axis, the
  $j$-axis and the line $i+j=0$ is filled with ``1'', then after $t$
  steps the evolution causes the whole triangle
  $\{(i,j) : i,j \le 0,\ i+j \ge -t \}$ to be filled with ``1''.
\end{itemize}
The last observation can be visualized as a displacement, at speed 1,
of the ``front'' formed by the line $i+j=0$, with a simultaneous
displacement (here a trivial one), at speed 0, of the ``fronts'' formed
by the $i$-\ and $j$-axis.  This combined displacement produces a
growing triangle full of ``1''.

The same observations hold if full lines are replaced by finite
segments, except that, depending on the values of neighboring spins,
in each iteration each segment can lose one or both of the ``1''
at its endpoints.  We conclude that if at some time the spider
\begin{eqnarray}
 &&{\rm SP}_{(0,0),L} =
\Bigl\{(i,0)\in\zed ^2 : -8L\le i \le 4L\Bigr\}\,\bigcup \,\nonumber\\
 &&\Bigl\{(0,j)\in\zed ^2 : -8L\le j \le 4L\Bigr\} 
 \bigcup\, \Bigl\{(i,j)\in\zed ^2 : i+j=0 \,,\,-6L\le i \le 6L \Bigr\}
  \label{eq:r.4}
\end{eqnarray}
is filled with ``1'', then after $4L$ iterations the ``1'' fill a
triangular region that contains the sphere $S_{(-L,-L),L}$.
Therefore, if $\mu$ is a invariant measure,
\begin{equation}
  \label{eq:r.5}
  \mu(1_{S_{(-L,-L),L}}) \;\ge\;
  \mu(1_{{\rm SP}_{(0,0),L}}) 
\;\ge\;\varepsilon^{3(12L+1)}\;.
\end{equation}
We have denoted $1_\Lambda$, for $\Lambda\subset\zed^{\!2}$, the event
$\{x:x(i,j)=1, (i,j)\in\Lambda\}$.  The last inequality in
\reff{eq:r.5} follows from the fact that a ``1'' has a probability at
least $\varepsilon$ to appear at a given site because of the noise.
As commented in Section \ref{s.ent}, such a probability is too large
for the invariant measure to be Gibbsian, or block-transformed
Gibbsian.

\paragraph{Example 2: North-South maximum of minima (NSMM)}.
The initial deterministic prescription is defined by
\begin{equation}
  \label{eq:r.6}
  x^{t+1}_{\rm det}(0,0) \;=\; 
\;=\; \max\Bigl\{\min\Bigl(x^t(0,0)\,,\, x^t(1,0)\Bigr)
\,,\, \min\Bigl(x^t(0,1)\,,\, x^t(1,1)\Bigr)\Bigr\}
\end{equation}
plus translation-invariance.  The corresponding evolution is not
symmetric under flipping, unlike the previous example.  The stochastic
version is obtained by adding one-sided noise as in \reff{eq:r1}.  For
small $\epsi$ this automaton has more than one invariant measure (see
lemma \ref{l.nonuniq}). One of them is, of course, the ``all-ones''
delta-measure $\delta_1$.

The mechanism for non-Gibbsianness for this model is even simpler to
describe than for the NEC model.  Indeed, it suffices to observe that
whenever a horizontal line is filled with ``1'', then in the next
iteration these ``1'' survive and in addition the parallel line
immediately to the South becomes also filled with ``1''.  The same
phenomenon happens for finite horizontal segments, except that each
creation of a new segment filled with ``1'' can be accompanied by
shrinkages of up to two sites (the spins at the endpoints) of all the
previously created segments.  We conclude that if the ``spider''
(which looks more like a snake in this case)
\begin{equation}
\widetilde{\rm SP}_{(0,0),L} \;=\;
\Bigl\{(i,0)\in\zed ^2 : -3L\le i \le 3L\Bigr\}
  \label{eq:r.7}
\end{equation}
is filled with ``1'' at some instant, then $2L$ instants later the
``1'' will cover at least a square region that includes the sphere
$S_{(0,-L),L}$. Arguing as for \reff{eq:r.5}, we obtain for all
invariant measures $\mu$ the bound
\begin{equation}
  \label{eq:r.8}
  \mu(1_{S_{(0,-L),L}}) \;\ge\;
  \mu(1_{\widetilde{\rm SP}_{(0,0),L}}) 
\;\ge\;\varepsilon^{6L+1}\;,
\end{equation}
which implies that $\mu$ is neither Gibbsian nor block-transformed
Gibbsian.
\medskip

A comment by A.~van~Enter (private communication) gives a colorful
description of the mechanism acting in both preceding examples: 
``the spider fills his stomach faster ($\asymp L$ sites at a time) 
than his legs shrink ($\asymp 1$ sites at a time)''.

\paragraph{Example 3: A non-example}.
The automata defined by the deterministic prescription
\begin{eqnarray}
 && x^{t+1}_{\rm det}(0,0) \;=\; \nonumber \\
 && \major\Bigl\{
 \min\Bigl(x^t(0, 2),\, x^t(-1, 2)\Bigr),\,
 \min\Bigl(x^t(2, 0),\, x^t( 2,-1)\Bigr),\,
 \min\Bigl(x^t(0,-1),\, x^t(-1, 0)\Bigr)\Bigr\}\nonumber\\
 \label{eq:r8}
\end{eqnarray}
followed by one-sided noise \reff{eq:r1}, also has multiple
invariant measures; this follows from lemma \reff{l.nonuniq}
(see below) because its $\sigma_0$ is empty.  Nevertheless,
neither the mechanism of Example 1 (travelling fronts),
nor that of Example 2 (growing strips) are present,
so the theory of the present paper does not apply.

\section{Non-nullness and the probability of aligned spheres}
\label{s.ent}

We present in this section the key property used in our paper to
detect non-Gibbsianness.  To state it in its natural generality we
introduce some definitions.

We consider a general space of the form $\Omega = S^{\zedd}$
where $S$ is some finite set, equipped with the usual product 
$\sigma$-algebra.  For $\Lambda\subset\zedd$ and $z\in\Omega$ 
we denote $z_\Lambda$ the cylinder\ 
\begin{equation}
  \label{eq:r40}
  z_\Lambda \;=\; \{x\in\Omega: x_i=z_i,
  i\in\Lambda\}\;. 
\end{equation}

\begin{definition}\label{d.asp}
A measure $\mu$ in $\Omega$ is said to have the {\bf
  alignment-suppression property (ASP)} if for every configuration
  $z\in\Omega$ 
\begin{equation}
  -\ln \mu(z_\Lambda) \succ |\Lambda| 
\label{eq:r45}
\end{equation}
for every finite set $\Lambda\subset\zedd$.
\end{definition}

Here and in the sequel $f \prec g$ or $g \succ f$, for $f$ and $g$
positive functions means that there exists a constant $C>0$ such that
$f \ge C g$.

All Gibbs measures have the ASP property, but many non-Gibbsian
measures too.  We construct now a general class of measures with this
property by considering renormalized measures having suitable
non-nullness features.  For this we consider an auxiliary
configuration space $\Omega_0 = S^{\zedd}$.  The single-site space
$S$ can be very general, not necessarily finite or even compact.  We
assume that there is a $\sigma$-algebra on $S$ and consider the usual
product Borel $\sigma$-algebra on $\Omega_0$.  A \emph{renormalization
  transformation} from $\Omega_0$ to $\Omega$ is a probability kernel
$T(\,\cdot\,|\,\cdot\,)$ from $\Omega_0$ to $\Omega$.  More
explicitly, for each $\omega \in \Omega_0$,~~ $T(\,\cdot\,|\omega)$ is
a probability measure in $\Omega$, and for each measurable event $A$
of $\Omega$, $T(A|\,\cdot\,)$ is a measurable function on $\Omega_0$.
In words, $T(A|\omega)$ is the probability that, given a configuration
$\omega \in \Omega_0$, the ``renormalized'' configuration is in $A$.
This represents a general stochastic transformation while
deterministic transformations are the special cases obtained via
delta-like prescriptions $T(\,\cdot\,|\omega)$.  A
\emph{block-renormalization transformation} is a transformation, for
which probabilities factorize in the following sense: to every $i \in
\zedd$ there corresponds a finite set $B(i)\subset\zedd$, called
\emph{block}, with the following properties:
\begin{itemize}
\item[(i)] If two points are far enough from each other, the
corresponding blocks are disjoint. That is, there is a
positive $d_0$ such that if the distance between
$k,\ell \in \zedd$ is greater than $d_0$, then
$B(k) \cap B(\ell) = \emptyset$ ($d_0=1$ for 
the renormalization transformations used in statistical mechanics,
while $d_0>1$ for common cellular-automata transformations).
\item[(ii)] If $i_1, \ldots, i_k$ are sites in $\zedd$, and
  $a_1,\ldots,a_k$ are values in ${\cal A}$, then
\begin{equation}
  T\Bigl(\{x_{i_1}=a_1,\ldots,x_{i_k}=a_k\}\Bigm|\, \omega\Bigr) \;=\;
  \prod_{j=1}^k \widehat T_{i_j}\Bigl(\{x_{i_j}=a_j\}\,\Bigm | \,
  \omega_{B(i_j)}\Bigr)\;. 
  \label{eq:r11}
\end{equation}
\end{itemize}
Our notation indicates that the functions $\widehat
T_{i_j}(\{x_{i_j}=a_j\}|\,\cdot\,)$ depend only on the values of
$\omega_{\ell}$ for $\ell\in B(i_j)$ (i.e., they are measurable with
respect to the $\sigma$-algebra generated by the cylinders with base
in $B(i_j)$).  Examples of such transformations include decimation,
(deterministic), Kadanoff transformations (stochastic), majority rule,
sign fields and transitions of cellular automata (the last three can
be deterministic or stochastic, depending on the setting).

The kernel $T$ naturally induces a transformation at the level of
measures: Each probability measure $\rho$ on $\Omega_0$ is mapped into
a probability measure $\rho T$ on $\Omega$ ---the \emph{renormalized
  measure}--- defined by
\begin{equation}
 \int_\Omega f(x)\, (\rho T)(dx) \;=\;
 \int_{\Omega_0} \,\Bigl[\int_\Omega f(x)\, T(dx|\omega)\Bigr]\,
 \rho(d\omega)\;,
 \label{eq:r12}
\end{equation}
for all suitable $f$ (e.g. continuous or non-negative measurable).
For each measure $\rho$ on $\Omega_0$ and each block $B(i)$
let us consider the conditional probabilities
$\rho(d\omega_{B(i)}\,|\,\omega_{\zedd\setminus B(i)})$.
For a given transformation $T$ we single out the set ${\cal P}_T$ of
measures on $\Omega_0$ that admit conditional probabilities such that
\begin{equation}
  \label{eq:r14}
\min_{a\in{\cal A}}\;\inf_{i\in\zedd}\;
\inf_{\;\omega_{\zedd\setminus B(i)}}\,
  \int \widehat T\Bigl(\{x_i=a\}\,\Bigm|\,\omega_{B(i)}\Bigr) \cdot
\rho\Bigl(d\omega_{B(i)}\,\Bigm|\,\omega_{\zedd \setminus B(i)}
\Bigr) \;\ge\; \delta \;,
\end{equation}
for some $\delta >0$.  We denote ${\cal P}$ the union of these
families ${\cal P}_T$ over all block-renormalization transformations
$T$.  Here is our key characterization.

\begin{theorem}\label{t.key}
Every measure in ${\cal P}$ has the alignment-suppression property.
\end{theorem}

\noindent
{\bf Proof.} Let $T$, $\rho$ be such that $\mu=\rho T$.  By property
(ii) above, there exists a constant $\gamma > 0$ (proportional to
$d_0$) such that for any $\Lambda\subset\zedd$ there is a family of
sites $i_1,\ldots,i_k \in \Lambda$ with $k \ge \gamma |\Lambda|$,
all of which are far enough from each other and therefore the 
blocks $B(i_1), \ldots,B(i_k)$ are disjoint.  We therefore have
that for every $z\in\Omega$
\begin{eqnarray}
  \mu(z_\Lambda) &=& \int 
\rho\Big(\widehat T\bigl(\{x_{i_1}=z_{i_1}\}\bigm|\,\cdot\,\bigr)
\,\Bigm|\,\omega_{\zedd\setminus B(i_1)} \Bigr)\,
\prod_{j=2}^k \widehat T_{i_j}\bigl(\{x_{i_j}=z_{i_j}\}\bigm | 
\omega_{B(i_j)}\bigr)\,\rho(d\omega)\nonumber\\
&\le & (1-\delta)\, 
\int \prod_{j=2}^k \widehat T_{i_j}\bigl(\{x_{i_j}=z_{i_j}\}\bigm | 
\omega_{B(i_j)}\bigr)\,\rho(d\omega)
  \label{eq:r17}\;.
\end{eqnarray}
This inequality is an immediate consequence of condition
\reff{eq:r14}.  After $k$ iterations of this procedure we obtain
\begin{equation}
  \label{eq:r19}
  \mu(z_\Lambda)  \;\le\; (1-\delta)^k
\;\le\; (1-\delta)^{\gamma |\Lambda|}\;. \QED
\end{equation}
\medskip 

The class ${\cal P}$ of measures is a very large class.  It contains
practically all block transformations of Gibbs measures with finite
alphabet obtained via standard statistical mechanics prescriptions
(decimation, Kadanoff, majority rule, etc), plus the measures
generated by finite-time evolutions of usual cellular automata
prescriptions.  There is by now a vast literature about such measures
---see, for instance, \cite{vEFS_JSP,lormae97,brikuplef98}; for recent
reviews with many references see \cite{ent98,fer98,fer99,entetal00}---
showing that many of them are non-Gibbsian.  In fact, the family
${\cal P}_I$, where $I$ is the identity, includes all \emph{uniformly
  non-null measures}.  These are measures $\mu$ that have, for each
finite region $\Lambda\subset\zedd$, uniformly bounded conditional
probabilities $\mu(d\omega_\Lambda\,|\,\omega_{\zedd\setminus
  \Lambda})$, that is, such that there exist $\delta_\Lambda>0$ with
\begin{equation}
  \label{eq:r30}
  \min_{a_\Lambda \in {\cal A}^\Lambda}\;
\inf_{\;\omega_{\zedd\setminus \Lambda}}\,
 \mu\Bigl(\{x_\Lambda=a_\Lambda\}\,\Bigm|\,\omega_{\zedd
   \Lambda} \Bigr) \;\ge\; \delta_\Lambda \;.
\end{equation}
We have denoted $a_\Lambda=(a_i)_{i\in\Lambda}$.  Gibbs measures are
uniformly non-null ---and in addition quasilocal (the finite-volume
conditional probabilities are continuous functions of the external
conditions $\omega_{\zedd \Lambda}$)--- hence they also belong to
${\cal P}_I$.  Property \reff{eq:r14} seems to be more general than
usual non-nullness, in particular it does not depend on the existence 
of a whole system of conditional probabilities.

The invariant measures of the automata of the present paper, on the
other hand, do not have the alingment-suppression property, hence 
they do not belong to the class ${\cal P}$.  They therefore can be 
neither Gibbsian nor uniformly non-null nor block-transformed Gibbsian.  
As further examples of measures without the ASP we mention the 
invariant measures of the basic voter model \cite{lebsch88}, 
the invariant measure of some non-local dynamics \cite{marsco91}, 
and the sign-fields of massless Gaussians \cite{lebmae87,dorvan89}, 
anharmonic crystals \cite[Section 4.4]{vEFS_JSP} and 
SOS models \cite{entshl98,lor98}.

For measures $\mu$ having a well defined relative entropy density
$s(\,\cdot\,|\mu)$, the alignment-supression property \reff{eq:r45}
implies that $s(\delta_z|\mu)>0$ for every periodic configuration
$z\in\Omega$.  The relative entropy density is known to exist for
translation-invariant Gibbs measures \cite[Chapter 15]{geo88}.  Recent
work in \cite{pfi00} shows that it is also well defined for most
translation-invariant measures obtained through block transformations
of Gibbs measures.  Because of this, the non-Gibbsianness resulting
from the lack of ASP has often been interpreted as ``large deviations
probabilities that are too large'' for Gibbsianness.  The
non-Gibbsianness (non-nullness) criterion obtained by falsifying
Theorem \ref{t.key}, however, is a more general argument that needs
neither translation invariance of $\mu$ nor the existence of the
entropy density.

For completeness, we mention a further result obtained in
\cite{entlor96}.  

\begin{theorem}\label{t.el}
  Suppose $\mu$ is a measure in $\Omega$ such that (i) it violates the
  ASP property for some periodic configuration $z\in\Omega$, and (ii)
  it is not a product measure.  Then, for every single-site
  block-renormalization transformation $T$ (i.e.\ a transformation
  defined by blocks $B(i)$ formed by only one site), the measure $\mu
  T$ is not Gibbsian.
\end{theorem}

This result follows from the fact that such a violation implies that
$s(\delta_z|\mu)=0$, which in turns implies that $s(\delta_zT|\mu
T)=0$.  If $\mu T$ were Gibbs, by a well known result \cite[Theorem
15.37]{geo88} the measure $\delta_z T$ would be Gibbs for an
equivalent interaction.  But this impossible because the latter is a
product measure and the former is not.  Note that if $T$ corresponds
to a not-totally asymmetric noise, the measure $\mu T$ is uniformly
non-null.  Hence its non-Gibbsianness would correspond to lack of
quasilocality.

For the automata of this paper, we suspect that many of its invariant
measures are non-product.  

\section{General Results}\label{s.gen}

We now describe a large family of automata exhibiting a general
version of the non-Gibbsianness mechanism of the first
two examples in Section \ref{s.simple}.
{}
Throughout the article we consider the $d$-dimensional integer space
$\zedd$ with $d>1$ embedded into the $d$-dimensional real space
$\IR ^d$ with the same axes and Euclidean norm $\|\cdot\|$.  
The configuration space is $\Omega = \SET{0,1}^{\zedd}$.
We first need some definitions. 

For any $i \in \zedd$ we denote $\tau_i~:~ \Omega \to \Omega$ 
the translation of $\Omega$ defined by $(\tau_i \; x)_j = x_{j-i}$.
Any function $f ~:~ \Omega \to \SET{0,1}$ will be called a 
transition function. Given any transition function $f$, we define the 
corresponding operator $D_{f} ~:~ \Omega \to \Omega$ by the rule 
\FORM{
  \forall~ i \in \zedd ~:~ (D_{f} \;  x)_i = f(\tau_i \; x).     \LL{D}
}

We call $f ~:~ \Omega \to \SET{0,1}$ {\em standard}
if it has the following three properties:
\begin{itemize}
\item[1)] $f$ is {\em local}, i.e. there is a finite set $\Delta
  \subset \zedd$ ---the \emph{support} of $f$--- such that
  $f(x) \equiv f(x_{\Delta})$.  Given $\Delta$, we denote $\rho$
  the maximum of $\|i\|$ for $i \in \Delta$.

\item[2)] $f$ is {\em monotonic}, that is
$
  (\forall ~i ~:~ x_i \leq y_i ) \IMPLIES f(x) \leq f(y). 
$ 

\item[3)] $f$ is not a constant. (Otherwise our theorem is either 
trivially true if $f \equiv 1$ or trivially false if $f \equiv 0$.)
\end{itemize}

Since $f$ is monotonic and non-constant, 
\FORM{
  f(\hbox{``all zeros''})=0 \HB{and}
  f(\hbox{``all ones''})=1.              \LL{non-const}
}

Let $\M$ denote the set of probability measures on $\Omega$ (on the
$\sigma$-algebra generated by cylinder sets).  For any $\epsi \in
[0,1]$ we define one-sided noise $N_{\epsi} : \M \to \M$ as follows:
when applied to a measure $\delta_x$ concentrated in a configuration
$x = (x_i)$, it produces a product measure $N_{\epsi} \; \delta_x$, in
which the $i$-th component equals 1 with a probability 1 if $x_i=1$
and with a probability $\epsi$ if $x_i=0$.

For any $x \in \Omega$ we denote its indicator
$\ind(x) = \SET{i \in \zedd ~|~ x_i =1}$.
Conversely, for any $S \subset \zedd$ we denote
$\conf(S)$ that configuration, whose indicator is $S$.

Let us call an element of $\IR ^d$ a {\em direction} if its norm
equals 1.  For any direction $p$ we call a {\em front} with this
direction any configuration whose indicator has the form
\FORM{ \SET{i
    \in \zedd \BAR \scalar{i,p} \leq C}, \LL{front}
}
where $C$ is a real number and $\scalar{\cdot,\cdot}$ denotes
the scalar product in $\IR ^d$. It is evident that
for any standard $f$ the operator $D_f$ transforms any front
(\RF{front}) into a front with the same direction, $C$ being
substituted by $C + V_p$, where $V_p$ does not depend on $C$.
We call $V_p$ the {\em velocity} of $D_f$ in the direction $p$.

Let us call a configuration $x \in \Omega$ invariant for $D_f$ 
if $D_f \: x = x$. Given $x,~ y \in \Omega$, we call $y$
a finite deviation of $x$ if the set of those $i \in \zedd$
for which $y_i \neq x_i$ is finite. We say that an invariant
configuration $x$ attracts $D_f$ if for any its finite
deviation $y$ there is a time $t$ such that $D_f^{\ t} \: y = x$.

\begin{theorem}\label{t.non}
Take any standard $f$, such that ``{all ones}'' attracts $D_f$, 
and make any one of the following two assumptions:\\
a) $V_p + V_{-p} \;\geq\; 0$ for all directions $p$.\\
b) There is a direction $p$ such that $V_p + V_{-p} > 0$.\\
Then for any $\epsi > 0$ all the invariant measures 
of $N_{\epsi} \: D_f$ satisfy
\FORM{
  -\ln \mu(\one_{S_{0,L}}) \prec L^{d-1}\;.                \LL{cylinder}
}
\end{theorem}

If $\epsi = 0$, our theorem may be false, for example if $D$ is 
the identity. Notice also that in the case b) our assumption that
``{all ones}'' attracts $D_f$ is redundant because it follows from b).

Let us present some further considerations that clarify the statement
of the theorem.  Given any non-constant affine function $\phi :
\IR ^d \to \IR $ and two numbers $C_1 \leq C_2$, we call a
{\em layer} any configuration $\conf\SET{i \in \zedd \BAR C_1
  \leq \phi(i) \leq C_2}$.  We call the thickness of this layer the
distance between the hyperplanes $\phi = C_1$ and $\phi = C_2$, that
is $(C_2 - C_1)/|\phi|$, where $|\cdot|$ is the norm.  We call a 
layer thick-enough if its thickness is not less than $2\rho$.

We call the two normal unit vectors to hyperplanes $\phi = \CONST$
the directions of this layer.  If $f$ is standard, $D_f$ transforms
any thick-enough layer into a layer with the same directions, the
thickness of the layer changing by $V_p + V_{-p}$. The condition a)
of our theorem means that thickness of any thick-enough layer does
not decrease and the condition b) means that thickness of some layer
increases under the action of $D_f$.

Of the examples of Section \ref{s.simple}, the NEC automaton satisfies
condition a), while the NSMM automaton satisfies condition b) for
$p=(0,1)$.  For the non-example, however, $V_p + V_{-p} < 0$ for all
directions $p$.  In all the three cases $f$ [given, respectively, by
\reff{r.2}, \reff{eq:r.6} and \reff{eq:r8}] is standard, and both
``{all zeros}'' and ``{all ones}'' attract $D_f$.

The NEC example is representative of a class of models with a
further duality property.  For any $x_i \in \SET{0,1}$ we denote
$\NEG{x_i} = 1-x_i$.  Accordingly, if $x$ is a configuration,
$\NEG{x}$ is another configuration such that
$(\NEG{x})_i \equiv \NEG{(x_i)}$.  Any transition function $f$
has an associated spin-flip function denoted $\NEG{f}$ and defined
by the identity $\NEG{f}(x) \equiv f(\NEG{x})$.
\footnote{In the theory of Boolean functions $\NEG{f}$
is called dual, but in the theory of random processes
the word ``duality'' is used for another purpose.}
Let us call $f$ self-spin-flip if it coincides with its spin-flip.
If $f$ is standard and self-spin-flip, then $V_p + V_{-p} \equiv 0$,
so the thickness of all layers does not change under the action of $D_f$.
For example, the function $\major(\cdot)$, described above, is self-spin-flip.
\medskip

It is evident that under the hypothesis of Theorem \ref{t.non}, the
measure $\delta_1$ is invariant for any superposition  
$N_{\epsi} \, D_f$ . Hence, the theorem is not trivial only if the
automata have more than one invariant measure. This is ensured by the
following lemma.

Given $f$, let us call a set $S \subset \zedd$ a one-set if
$f(\conf(S))=1$.  Since one-sets belong to $\zedd$, they belong
to $\IR ^d$, where we can consider their convex hulls, the
intersection of which is denoted $\sigma_1$.  {} In the analogous way
we call a set $S \subset \zedd$ a zero-set if
$f(\conf(\zedd - S))=0$ and denote $\sigma_0$ the intersection
of their convex hulls.

\begin{theorem}\label{l.nonuniq}
For any operator $D_f$ defined by (\RF{D}), where $f$ is standard, 
the following four statements are equivalent:
\begin{itemize}
\item[1)] $N_{\epsi} \: D_f$ has more than one invariant measure for
  some positive $\epsi$.
\item[2)] The configuration ``{all zeros}'' attracts $D_f$.
\item[3)] $\sigma_0$ is empty.
\item[4)] There are a natural number $m \leq d+1$ and $m$ affine functions\\
  $\phi_1,\ldots,\phi_m~:~\IR ^d \to \IR $ such that:
\[
\left\{
\begin{minipage}{6.6in}
\begin{itemize}
\item[i)] for every $j \in [1,m]$ the set $\SET{p \in \zedd ~:~
    \phi_j(p) \leq 0}$ is a zero-set.
\item[ii)] $\phi_1 + \cdots + \phi_m \equiv \CONST > 0$.
\item[iii)] There is a rational point $p \in \IR ^d$ such that
  $\phi_j(p) > 0$ for all $j \in [1,m]$.
\end{itemize}
\end{minipage}
\right.
\]
\end{itemize}
\end{theorem}

\section{Proof of theorem \protect\ref{l.nonuniq}}

If we omit the condition iii) in 4), our theorem \ref{l.nonuniq}
almost follows from theorems 5 and 6 and lemma 12 of \cite{too80}.
However, there is some difference, so for the reader's convenience we
completely deduce 4) from 3).  

Suppose that $\sigma_0$ is empty.  Every zero-set can be represented
as an intersection of several zero-half-spaces, i.e.\ half-spaces,
which are zero-sets, where a half-space is a subset of $\IR ^d$,
where some non-constant affine function does not exceed zero.  Thus
there are several zero-half-spaces, whose intersection is empty.
Everyone of them can be represented as $\SET{p \in \IR ^d \BAR
  f_i(p) \leq 0}$, where $f_i$ are affine functions on $\IR ^d$.
We can choose these functions so that they have no common
direction of recession (that is, no direction $p$ such that
$f_i(p) \le f_i(0)$ for all $i$), which allows us to
apply to them Theorem 21.3 on page 189 of \cite{roc70}.
Since the intersection of our zero-half-spaces is empty, 
the case (a) of this theorem is excluded in the present
situation, whence the case (b) takes place, which amounts 
to our conditions i) and ii) in 4), the products $\lambda_i f_i$ 
mentioned in the case (b) serving as our $\phi_i$.
{}
We may assume that our $m$ is the minimal for which 
there are functions satisfying i) and ii). Based on 
this, let us prove statement iii) using the following
lemma, which is a direct consequence of
Theorem 21.1 on page 186 of \cite{roc70}:

\begin{lemma} \LL{Theorem 21.1}

Let $\phi_1,\ldots,\phi_m$ be affine functions on $R^d$.
Then one and only one of the following alternatives holds:
\begin{itemize}
\item[(a)] There exists some $x \in \IR^d$ such that
$
  \phi_1(x) > 0,\ldots,\phi_m(x) > 0;
$
\item[(b)] There exist non-negative real numbers
  $\lambda_1,\ldots,\lambda_m$, not all zero, such that
  the sum $\lambda_1 \phi_1(x) + \cdots + \lambda_m \phi_m(x)$
  is a non-positive constant.
\end{itemize}
\end{lemma}

Let us assume that the case (b) takes place in our situation.
We may assume that $\lambda_m$ is the greatest of 
$\lambda_1,\ldots,\lambda_m$, and therefore positive. 
From the statement ii) of 4), not all $\lambda_i$ are 
equal to $\lambda_m$. Let us divide all terms by $\lambda_m$:
\[
  \frac{\lambda   _1 }{\lambda_m} \phi_1 + \cdots +
  \frac{\lambda_{m-1}}{\lambda_m} \phi_1 + \phi_m = \CONST \leq 0
\]
and subtract this from the statement ii) of 4):
\[
  \BRA{1 - \frac{\lambda   _1 }{\lambda_m}} \phi_1 + \cdots +
  \BRA{1 - \frac{\lambda_{m-1}}{\lambda_m}} \phi_1 = \CONST \geq 0.
\]
Here all coefficients are non-negative and not all are zero.
Therefore the functions $(1 - \lambda_i/\lambda_m) \phi_i$ for
$i=1,\ldots,m-1$ also satisfy the conditions i) and ii) of 4)
with a smaller value of $m$, which contradicts our assumption.
Thus case (b) is excluded, so case (a) takes place, whence
there is a point $p \in \IR ^d$ where all $\phi_j(p) > 0$.
Since all $\phi_j$ are continuous, there is a rational point
with this property also, whence condition iii) of 4) follows. $\qed$

\section{Proof of Theorem \protect\ref{t.non}}

\subsection{Proof of \protect(\RF{cylinder}) in case a) of the theorem}

Rewording Lemma \ref{l.nonuniq} for the case when 0 and 1 are permuted,
we see that whenever $f$ is standard and ``{all ones}'' attracts
$D_f$, there exist a natural number $m \leq d+1$ and $m$ affine
functions $\phi_1,\ldots,\phi_m~:~\IR ^d \to R$ such that:
\FORM{ \left.
\begin{minipage}{6in}
\begin{itemize}
\item[i)] for every $j \in [1,m]$ the set $\SET{i \in \zedd ~:~
    \phi_j(i) \leq 0}$ is a one-set.
\item[ii)] $\phi_1 + \cdots + \phi_m \equiv \CONST > 0$.
\item[iii)] There is a rational point $p \in \IR ^d$ 
such that $\phi_j(p) > 0$ for all $j \in [1,m]$.
\end{itemize}
\end{minipage}
~~\right\}                                               \LL{1}
}

For instance, for the NEC example there are $m=3$ such 
affine functions, whose level lines are horizontal, 
vertical and lines of slope $-1$ respectively.

For every $j$ let us denote $\PHI_j = \phi_j - \phi_j(0)$ whence$\phi_j
= \PHI_j + \phi_j(0)$, where $\PHI_j$ is the linear part.  Notice that
$|\phi_j(0)| \leq |\phi_j| \cdot \rho$ and that $\phi_1(0) + \cdots +
\phi_m(0) > 0$.  {} Notice also that if $f$ is standard, ``{all
  ones}'' attracts $D_f$ and $V_p + V_{-p} \geq 0$ for all directions
$p$, then for any $j \in [1,m]$ and any thick-enough layer $y =
\conf\SET{i \in \zedd \BAR C_1 \leq \phi_j(i) \leq C_2}$ \FORM{
  \ind(D_f \: y) \supseteq \Bigl\{i \in \zedd \BAR C_1 + \phi_j(0)
  \leq \phi_j(i) \leq C_2 + \phi_j(0)\Bigr\}.  \LL{layer} } (See
immediately above \reff{front} for the definition of Conf and Ind.)

\begin{lemma} \label{l.2}
Take any standard $f$ and assume that ``{all ones}'' attracts
$D_f$ and that $V_p + V_{-p} \geq 0$ for all directions $p$.
Take $x^*$ defined by 
\FORM{
  \ind(x^*) = \bigcup_{1 \leq j \leq m} 
  \Bigl\{i \in \zedd \BAR |\phi_j (i)| \leq 2\rho \cdot
  |\phi_j|\Bigr\}.   \LL{x} 
}
Then for $t=0,1,2,3,\ldots$ the indicator 
of $D^t \: x^*$ includes the union $A_t \cup B_t$, where
\FORM{
  A_t = \bigcup_{1 \leq j \leq m} \Bigl\{i \in \zedd \BAR  
  |\L_j(i) - t \cdot \phi_j(0)| \leq 2\rho \cdot |\phi_j|\Bigr\}
\LL{X} 
}
and
\FORM{
  B_t = \bigcap_{1 \leq j \leq m} \Bigl\{i \in \zedd \BAR 
   \PHI_j(i) - t \cdot \phi_j(0) \leq 0\Bigr\}.                       \LL{O}
}
\end{lemma}

[For the NEC example of Section \ref{s.simple}, this lemma corresponds
to observation (iv).]

Let us prove this lemma by induction.
{}
Base of induction: Since $A_0$ coincides with $\ind(x^*)$ 
and $B_0 \subset A_0$, our statement is true for $t=0$.

{\bf Induction step.~} 
Let us suppose that $\ind(D^t \; x^*) \supseteq A_t \cup B_t$,  
take any $i \in A_{t+1} \cup B_{t+1}$ and prove that 
$i \in \ind(D^{t+1} \, x^*)$. Let us consider two cases.

{\bf Case 1.~} Let $i$ belong to $A_{t+1}$.
Then our statement follows from (\RF{layer}).

{\bf Case 2.~} Let $i$ belong to $B_{t+1}$, but not to $A_{t+1}$. Then 
\[
  \PHI_j(i) - (t+1) \cdot \phi_j(0) \leq - 2\rho \cdot |\phi_j|
\]
for all $j \in [1,m]$. Notice that
\[
  \PHI_j(i + v_k) \leq 
  \PHI_j(i) + |\phi_j| \cdot |v_k| \leq \PHI_j(i) + |\phi_j| \cdot \rho.
\]
Therefore
\[
  \PHI_j(i + v_k) - t \cdot \phi_j(0) \leq 
  \PHI_j(i) + |\phi_j| \cdot \rho - (t+1) \phi_j(0) + \phi_j(0) \leq
  - 2 \rho \cdot |\phi_j| + |\phi_j| \cdot \rho + \phi_j(0) \leq 0.
\]
Thus
\[
  i + \Delta \subset B_t \subset \ind(D^t \: x^*).
\]
Hence from (\RF{non-const}) $i \in \ind(D^{t+1} \: x^*)$. Lemma
\ref{l.2} is proved.

\begin{lemma}\label{l.3}
Under the hypotheses of Lemma \ref{l.2}, there is a positive constant 
$\alpha > 0$ such that for all $t=0,1,2,\ldots$ the set 
$B_t$ defined by (\RF{O}) contains a sphere in $\zedd$ 
with the radius $\alpha \cdot t$. 
\end{lemma}

{\bf Proof.~} In fact we shall prove that
\[
  \forall~ i \in \zedd,~ t=0,1,2,\ldots ~:~ 
  |i + t \cdot p| \leq \alpha \cdot t \IMPLIES i \in B_t,
\]
where $p$ is that rational point where all $\phi_j(p) > 0$, whose 
existence is provided by iii). Let us denote $\kappa_j = \phi_j(p) > 0$ 
and $\alpha = \min_j (\kappa_j/|\phi_j|)$, that is the minimal distance 
from $p$ to the hyperplanes $\phi_j = 0$. Let us consider three cases.

{\bf Case 1:~} $p=0$. Then $\phi_j(0) = \kappa_j > 0$ for all $j$.
Now let us take any point $i$ in the sphere with the radius 
$\alpha \cdot t$ and center $0$. This means that
\[
  |i| \leq \alpha \cdot t = \min_j(\phi_j(0)/|\phi_j|) \cdot t.
\]
Then
\[
  \PHI_j(i) \leq |i| \cdot |\phi_j| \leq 
  \phi_j(0)/|\phi_j| \cdot t \cdot |\phi_j| = t \cdot \phi_j(0)
\]
for all $j$, whence $i \in B_t$.

{\bf Case 2:~} $p \in \zedd$. Then along with our
operator $D_f$ we consider another operator $D_g$, where
$g(x) \equiv f(\tau_{p} \: x)$. The function $g$ is also standard,
$D_g$ is also attracted by ``{all ones}'' and the affine functions
provided for $D_g$ by iii) of (\RF{1}) can be obtained from those
for $D_f$ by the same translation, so their values at 0 are
$\kappa_1,\ldots,\kappa_m > 0$, whence $D_g$ fits our case 1.
So the set $B_t$ for $D_g$ contains a sphere with the center 0
and radius $\alpha \cdot t$. Since $D_f$ commutates with all translations,
the set $B^t$ for $D_g$ results from the set $B_t$ for $D_f$
by a translation at $t \cdot p$. Thus the set $B_t$ for $D_f$
results from $B_t$ for $D_g$ by the opposite translation, whence it 
contains a sphere with the center $- t \cdot p$ and the same radius.

{\bf Case 3:~} $p$ is any rational point.  Let us denote $q$ the least
common denominator of all the coordinates of $p$ and immerse our
$\zedd$ into the set $\zed _q^d$, where $\zed _q =
\SET{n/q \BAR n \in \zed }$.  Let us denote $\Omega_q =
\SET{0,1}^{\zed _q^d}$.  Now $f$ can be considered as a function
$g$ from $\Omega_q$ to $\SET{0,1}$.  Now let us ``stretch''
$\zed _q^d$ to turn it into $\zedd$.  Under this stretch
the function $g$ remains standard and ``{all ones}'' still attracts
$D_g$. In addition to that, the affine functions for $D_g$ with the
properties (\RF{1}) now can be obtained from those for $D_f$ by a
homothety with coefficient $q$.  Therefore their values at the integer
point $q \cdot p$ are $\kappa_1,\ldots,\kappa_m > 0$. So $D_g$ fits
our case 2, whence the set $B_t$ for $D_g$ contains a sphere with the
center $-t \cdot q \cdot p$ and radius $\alpha \cdot q \cdot t$,
whence the set $B_t$ for $D_f$ contains a sphere with the center $-t
\cdot p$ and radius $\alpha \cdot t$. Lemma \ref{l.3} is proved.
$\qed$ \medskip

Now let us prove (\RF{cylinder}). From monotonicity it is sufficient 
to prove this inequality for $\mu = (N_{\epsi} \, D_f)^t \; \delta_0$ for 
some $t$. Let us choose $t_1$ such that $\alpha \cdot t_1 \geq R + d$.
Then, taking $x^*$ defined by (\RF{x}) 
as the initial configuration, after $t_1$ 
time-steps we obtain a configuration, whose indicator contains a
sphere with the radius $R + d$ and therefore contains a sphere 
with the radius $R$ and center at some integer point $p$.
{}
However, what we actually need is a finite deviation from
``{all zeros}'', which coincides with $x^*$ only within a sphere 
with the radius $R + t_1 \cdot \rho$ and has zeros outside it. 
The cardinality of its indicator does not 
exceed $C(R^{d-1} + 1)$ with an appropriate $C$.
{}
Translating this configuration at the vector $-p$, we obtain
another configuration, which fills with ones a sphere with 
radius $R$ and center at the origin after $t_1$ time-steps.
{}
The probability that the actual configuration's indicator
contains this configuration is not less that $\epsi^{C (R^{d-1} + 1)}$,
whence (\RF{cylinder}) follows. $\qed$


\subsection{Proof of \protect(\RF{cylinder}) in case b) of the theorem}

This time we define $x^*$ as follows:
\[
  \ind(x^*) = \Bigl\{i \in \zedd \BAR |\scalar{i,p}| \leq
  \rho\Bigr\}.   
\]
Then for all $t=0,1,2,\ldots$
\[
  \ind(D_f^t \: x^*) \;\supseteq\; \Bigl\{i \in \zedd \BAR 
  -\rho + t \cdot V_{-p} \leq \scalar{i,p}| \leq \rho + t \cdot
  V_p\Bigr\}\;.   
\]
Here the right side is a layer with the thickness $2\rho + t(V_p + V_{-p})$.
Given any $R \geq 0$, let us choose the minimal integer $t_1$ 
for which $2\rho + t_1 (V_p + V_{-p}) \geq R + d$. 
Then indicator of $D_f^{t_1} \: x^*$ contains a sphere with 
an integer center and radius $R$. If we take an initial
condition which coincides with $x^*$ within a sphere with
the center at the origin and radius $R + d + t_1 \cdot \rho$,
we shall obtain the same result. This configuration has 
$C(R^{d-1} + 1)$ components that equal 1, where $C$ is an 
appropriate constant. Further we argue like in case a). $\qed$

\section{Final notes}

{\bf Note 1.~} Using minoration arguments, is is easy to 
expand our theorem to some random cellular automata, 
which cannot be represenred as $N_{\epsi} \: D_f$. 
Using the same $\Delta$ as before and choosing 
transition probabilities $\theta(x | y_{\Delta})$ for all 
$x \in \SET{0,1}$ and $y \in \SET{0,1}^{\Delta}$, we can define
a random cellular automaton as an operator $P ~:~ \M \to \M$ 
which transforms any $\delta_y$, where $y \in \Omega$, into a 
product-measure in which the probability that the $i$-th
component equals $x$ is $\theta(x | y_{i+\Delta})$.
This operator majorates $N_{\epsi} \: D_f$ if
\[
  \theta(x | y_{\Delta}) 
  \cases{ =1 & if $f(y_{\Delta})=1$,\cr
  \geq \epsi & if $f(y_{\Delta})=0$.
       }
\]
As soon as this condition holds and $D_f$ satisfies conditions 
of our theorem, all invariant measures of $P$ also satisfy 
(\RF{cylinder}) and therefore are non-Gibbs.

{\bf Note 2.~} In some cases it is possible to obtain
a stronger estimation than (\RF{cylinder}).
Let $d > a > 0$ and $f(x)$ equal
\[
   \min_{i_{1  },\ldots,i_a \in \SET{0,1}} \quad 
   \max_{i_{a+1},\ldots,i_d \in \SET{0,1}} \quad 
   x(i_1,\ldots,i_d) 
\]
where $i_1,\ldots,i_d$ are the coordinates of $\zedd$.
In this case
\[
  - \ln \mu(\one(S_{0,L})) \prec L^a,
\]
where $\mu$ is any invariant measure of $N_{\epsi} D_f$.
If $a < d-1$, this estimation is stronger than (\RF{cylinder}).
This estimation can be proved in the same manner as in the case b),
only $x^*$ now is defined by the condition:
\[
  x^*_i=1 \HB{if} \max(|i_{a+1}|,\ldots,|i_d|) \leq \CONST.
\]

{\bf Note 3.~} Given a standard $f$, let us assume that
``all zeros'' attracts $D_f$. Then we hope to estimate
$-\ln \mu(\one(S_{0,L}))$ from below as follows:
\[
  -\ln \mu(\one(S_{0,L})) \succ L.
\]
If we succeed, this will settle the question of asymptotics of
$-\ln \mu(\one(S_{0,L}))$ in some cases, e.g. in our examples 1 and 2,

{\bf Note 4.~} Those conditions under which our theorem 
holds and is non-trivial can be satisfied only for $d > 1$.
However, a statement similar to our theorem for the
one-dimensional case was proved in \cite{too72}.
Namely, it was proved that all non-trivial invariant measures
of a class of one-dimensional random cellular automata did
not belong to a class, which included all Markov measures.

\section*{Acknowledgments}  We thank Aernout van Enter for very useful
comments and criticism.  We also thank the organizers and funding
agencies of the IV Brazilian School of Probability where the final
discussions for this paper took place.  RF wishes to acknowledge
travel support by an agreement USP-Cofecub (projet UC/68/99,
\emph{Comportement \`a temps interm\'ediaire des syst\`emes
  al\'eatoires}).

\label{LP}

\end{document}